\documentclass[
reprint,
amsmath,
amssymb,
aps,
prl,
twocolumn,
]{revtex4-2}

\usepackage{xcolor}
\usepackage{graphicx}
\usepackage{dcolumn}
\usepackage{bm}
\usepackage{float}

\begin{document}

\title{Ring-shaped atom-trap lattices using multipole dressing fields}

\author{Fabio Gentile}\thanks{These authors made equal contributions.}
\author{Jamie Johnson}\thanks{These authors made equal contributions.}
\author{Konstantinos Poulios}
\author{Thomas Fernholz}
\email{email: thomas.fernholz@nottingham.ac.uk}
\affiliation{School of Physics \& Astronomy, University of Nottingham, University Park, Nottingham NG7 2RD, U.K.}

\date{\today}

\begin{abstract}
We present a method for the creation of closed-loop lattices for ultra-cold atoms using dressed potentials. We analytically describe the generation of trap lattices that are state-dependent, with dynamically controlled lattice depths and positioning. In a design akin to a synchronous motor, the potentials arise from the combination of a static, ring-shaped quadrupole field and multipole radio-frequency fields.
Our technique relies solely on static and radio-frequency (rf) magnetic fields, enabling the creation of robust atom traps with simple control via rf amplitudes and phases.  
Potential applications of our scheme span the range from quantum many-body simulations to guided Sagnac interferometers.
\end{abstract}

\pacs{Valid PACS appear here}


\maketitle


Adiabatic radio-frequency (rf) dressed potentials play an increasingly important role in recent developments with ultra-cold atomic, see physics\cite{garraway_jphysb_2016} for a recent review.
These potentials, produced by the combination of static and oscillating magnetic fields, enable the creation of a plethora of different trapping geometries that can support confinement as well as, under certain conditions, transport of atomic clouds.

The trapping potentials obtained with rf dressing techniques are inherently species-selective and state-dependent, because they can be controlled independently for atoms that have different Land\`e g-factors \cite{lesanovsky_pra_2006, bentine_jphysB_2017}. For the case of often used alkali atoms, the different hyperfine levels of their electronic ground state have g-factors with near identical magnitude but opposite sign. Arbitrary superpositions of such internal states can be prepared, and the combination with dressed, state-dependent potentials allows for coherent beam splitting, using only fields oscillating in the rf and microwave (mw) regime.
This results in a versatile workbench capable of generation, independent manipulation, and detection of internally labelled atomic superposition states, especially relevant for interferometric schemes \cite{cronin_revmodphys_2009,stevenson_prl_2015}.
There are numerous examples of experimental implementations that demonstrate the generation and detection of dressed superposition states [see \cite{garraway_jphysb_2016} and references therein], including also non-destructive detection methods \cite{jammi_pra_2018}.
Recently, transport of a single, dressed spin state has been demonstrated over macroscopic distances \cite{pandey_nature_2019}.
By manipulating the amplitude, frequency, and polarization or relative phases of the contributing rf fields, versatile control over the resulting rf dressed potentials can be achieved \cite{fernholz_pra_2007, lesanovsky_prl_2007, harte_pra_2018}.
This ability to dynamically modify the potential landscape between different configurations within a single experimental run is significantly useful in the context of both fundamental as well as applied experiments with ultra-cold atomic clouds and Bose-Einstein condensates (BECs).
The possibilities range from double wells \cite{schumm_nphys_2005, bohi_nphys_2009} to hollow-shell traps \cite{colombe_epl_2004}, ring-shaped matter-waveguides \cite{heathcote_njp_2008, sherlock_pra_2011, navez_njp_2016, bell_pra_2018} and purely magnetic atom-trap lattices \cite{sinuco-leon_njp_2015, wang_sbull_2016}.

In recent years, ultra-cold atoms in optical, magnetic or hybrid trap lattices have constituted an important quantum simulation testbed for a variety of physical phenomena otherwise not straightforward to probe \cite{gross_science_2017}.
These include studying the dynamics of strongly correlated particles \cite{preiss_science_2015} and investigating new topological phases of matter \cite{lohse_nphys_2016,lohse_nature_2018} as well as the thermalization of quantum systems and dynamics in the many-body regime \cite{amico_prl_2005}. The use of dressed potentials may add to this, as magnetic atom-trap lattices with interesting topologies can be formed, including, e.g., ring structures, that are furthermore adjustable and dynamically controllable in a state dependent fashion.
These ring-shaped atom-trap lattices have been proposed as analogues for superconducting flux qubits \cite{amico_srep_2014} as well as platforms where artificial gauge fields can be studied \cite{victorin_pra_2018} and correlated many-body effects can be harnessed for the implementation of rotation sensors and gyroscopes with enhanced sensitivity \cite{naldesi_arxiv_2019}.
Here we present a method that allows for the generation of such ring-shaped atom-trap lattices based on rf dressed potentials, by combining a ring-shaped quadrupole potential with a multipole rf field.
Our method is compatible with atom-chip technology \cite{keil_jmodopt_2016} and may enable robust, mechanically stable and compact quantum devices and sensors.

Rf dressed potentials arise from the combination of an inhomogeneous magnetic field and an oscillating magnetic field that drives atomic spin flips. The static field defines a two-dimensional manifold where Larmor precession can be resonantly excited, thus coupling low-field seeking states to high-field seeking states. This principle can be extended to any pair of such states, e.g., by coupling states from different hyperfine manifolds, where the nuclear spin changes orientation with respect to the electronic spin. A trap is formed in the regime where atoms traverse the region of resonance adiabatically. 
The trap topology of our scheme is based on an axially symmetric combination of a ring-shaped, static quadrupole field and an oscillating rf field with radial and axial components of different phases. Such an arrangement results in a dressed magnetic potential with toroidal geometry~\cite{fernholz_pra_2007}, which allows for the creation of ring-shaped and toroidal, i.e.\ hollow torus-shaped atom traps. The fields are specifically chosen such that connected potential minima without degenerate points are generated, which would otherwise cause atom loss. 
In the following, we first recapitulate the approach for generating hollow-torus and in particular ring-shaped atom traps before describing the method for partitioning these traps in order to form a lattice.

An atom interacting with a weak magnetic field, consisting of a static term $\textbf B_\text{dc}=B_\text{dc}\mathbf{e}_0$ and a time-dependent term $\textbf{B}_\text{ac}(t)=\textbf{B}_\text{rf} e^{i\omega_\text{rf} t}/2+ c.c.,\, \textbf{B}_\text{rf} \in \mathbb{C}^3$ that oscillates at frequency $\omega_\text{rf}$, is described by the Hamiltonian
\begin{equation}
\hat{H}\left(t\right)=g_F\mu_B\frac{\hat{\mathbf{F}}}{\hbar}\cdot\left(\mathbf{B}_{\mathrm{dc}}+\mathbf{B}_{\mathrm{ac}}(t)\right),
\end{equation}
\noindent where $\hat{\mathbf{F}}$ is the atom's total angular momentum, $\mu_B$ is the Bohr magneton, and $g_F$ is Land\'e g-factor. In general, the static field direction depends on position, and the oscillating field can be expressed in a local spherical basis $\lbrace \mathbf{e}_0,\mathbf{e}_\pm=(-\mathbf{e}_1\pm i \mathbf{e}_2)/\sqrt{2}\rbrace$ as $\textbf{B}_\text{rf}=B_+\mathbf{e}_++B_-\mathbf{e}_-+B_0\mathbf{e}_0$, with amplitudes $B_{\pm,0}$ of corresponding field polarizations that drive $\sigma^{\pm}$- and $\pi$-polarised transitions with respect to a quantization axis $\mathbf{e}_0$. Vectors $\mathbf{e}_{0,1,2}$ form a right-handed system and we use corresponding, dimensionless spin operators $\hat{F}_{0,1,2}= \mathbf{e}_{0,1,2}\cdot\hat{\mathbf{F}}/\hbar$. 
The Hamiltonian can be transformed to a frame rotating at $\omega_\text{rf}$ (the rf dressing frequency) about the static field direction $\mathbf{e}_0$, i.e.\ $\hat{H}'=\hat{U}\hat{H}\hat{U}^{-1}+i\hbar(\frac{\partial}{\partial t}\hat{U})\hat{U}^{-1}$ with the unitary operator $\hat{U}=e^{i\omega_\text{rf}t\hat{F}_0 }$.
Defining $\hat{F}_\pm=\hat{F}_1\pm i\hat{F}_2$, using the Baker-Haussdorff formula $\hat{U}\hat{F}_\pm\hat{U}^{-1}=e^{\pm i\omega_\text{rf}t}\hat{F}_\pm$, and making the rotating wave approximation (RWA), leads to the transformed Hamiltonian
\begin{equation}
    \hat{H}'_\text{RWA}=\frac{1}{2}g_F\mu_B\left((B_\text{dc}-\frac{\hbar\omega_\text{rf}}{g_F\mu_B})\hat{F}_0-\frac{B_+}{\sqrt{2}}\hat{F}_-\right)+h.c.
\end{equation}
The resulting spectrum of dressed state (quasi)energies is given by
\begin{equation}\label{eq:EmF}
E_{m_F}=m_Fg_F\mu_B\sqrt{\left(B_{\text{dc}}-\frac{\hbar\omega_{\text{rf}}}{g_F\mu_B}\right)^2+\frac{|B_+|^2}{2}},
\end{equation}
\noindent where $m_F$ is the magnetic quantum number, and the amplitude of the remaining dressing field component is given by $B_+=\mathbf{e}_+\cdot\textbf{B}_\text{rf}$. 
The vanishing of the first term defines a resonance, which may occur for negative frequency $\omega_\text{rf}$ and thus inverted rotational senses, depending on the sign of $g_F$ and the chosen sign of $B_\text{dc}$.

\begin{figure}
\centering
\includegraphics[width=1 \columnwidth]{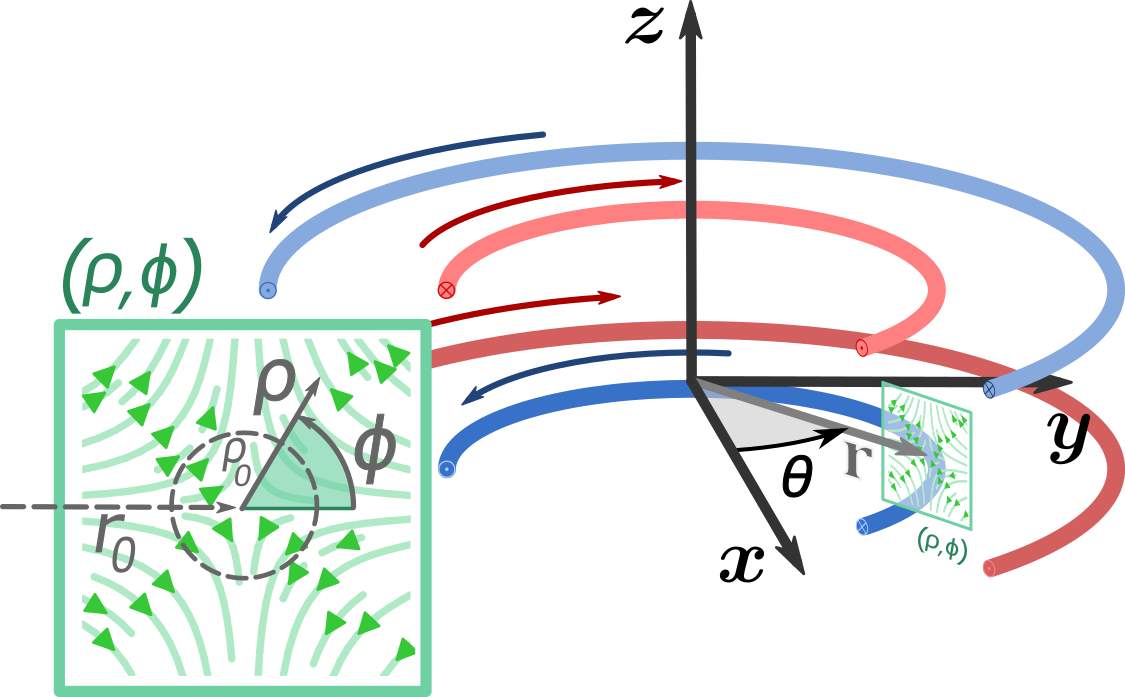}
\vspace{-0.3 cm}
\caption{\label{fig:ring-coordinates}Illustration of the static ring quadrupole field (section with field lines and local polar coordinates in the inset). Such a field can be obtained by means of four counterpropagating current loops as indicated by the arrows. Global Cartesian and cylindrical coordinates are shown together with local polar coordinates, defining toroidal ($\theta$) and poloidal ($\phi$) angles. \vspace{-0.3 cm} }
\end{figure}

We assume a static, circular quadrupole field with zero magnetic field along a ring of radius $r_0$, centered in the laboratory's $x,y$-plane at $z=0$, see~Fig.~\ref{fig:ring-coordinates}. Such a field can be generated using four counter-propagating circular currents. We approximate the static field in the vicinity of this ring, and use the toroidal angle $\theta$ together with local polar coordinates $\rho,\phi$ to parameterize planes orthogonal to the ring. The approximate field magnitude is then given by $B_{\text{dc}}=q\rho$, where $q$ is the quadrupole gradient and $\rho$ is the distance from the field zero. For $m_Fg_F>0$, the resonance condition $q\rho_0=\hbar\omega_{\text{rf}}/g_F\mu_B$ minimizes the potential with respect to $\rho$, thus defining the surface of a torus at $\rho=\rho_0$ where atoms can be trapped~\cite{fernholz_pra_2007}. The potential on this surface is given by the locally varying amplitude $B_+$, which is the focus of the remaining discussion. For its evaluation, we define a local coordinate system in the vicinity of the ring of zero field with a basis given by the approximated static field direction $\mathbf{e}_0$, the tangent to the ring $\mathbf{e}_1$, and the right-handed completion $\mathbf{e}_2=\mathbf{e}_0\times\mathbf{e}_1$.
Expressed in the Cartesian laboratory representation this choice of local basis is given by
\begin{eqnarray}
\mathbf{e}_0&=&(-\cos\theta \cos\phi, -\sin\theta \cos\phi, \sin\phi)^T, \label{eq:e0}\\
\mathbf{e}_1&=&(-\sin\theta, \cos\theta,0)^T,\\
\mathbf{e}_2&=&(-\cos\theta\sin\phi,-\sin\theta\sin\phi,-\cos\phi)^T.\label{eq:e2}
\end{eqnarray}

In order to generate non-vanishing potential minima, the effective field amplitude $B_+$ must be non-zero. One possibility to achieve this with axial symmetry is to use an rf field tangential to the ring, i.e.\ parallel to $\mathbf{e}_1$. For an alternating current $I$ along the setup's central (c) $z$-axis, the field would be $\mathbf{B}_\text{rf}^\text{(c)}=\mathbf{e}_1\mu_0 I/2r$ and lead to $B_+^\text{(c)}=\mathbf{e}_+\cdot\mathbf{B}_\text{rf}^\text{(c)}=-\mu_0 I/2\sqrt{2}r$. More versatility can be achieved by using a dressing field that is elliptically polarized in the $\rho$, $\phi$ planes.
Such a toroidal (t) field can be generated by combining a uniform field, linearly polarized along the $z$-direction, with a phase-shifted, axially symmetric quadrupole field that provides a radial component in the $x,y$-plane along $r$. For simplicity, we neglect radial dependence of this field and approximate it in the vicinity of the forming trap. In the Cartesian laboratory presentation, its decomposition into orthogonal circular components is given by
\begin{equation}\label{eq:B1}
\textbf{B}_\mathrm{rf}^\text{(t)}=
\frac{a_+}{\sqrt{2}}
\begin{pmatrix}
\cos\theta\\
\sin\theta\\
i
\end{pmatrix}
+
\frac{a_-}{\sqrt{2}}
\begin{pmatrix}
\cos\theta\\
\sin\theta\\
-i
\end{pmatrix},
\end{equation}
\noindent with amplitudes $a_+$ and $a_-$. 
In this case, the coupling field component is given by the projection
\begin{equation}\label{eq:betapm1}
B_+^\text{(t)}=\textbf{e}_+\cdot\mathbf{B}_\mathrm{rf}^\text{(t)}=\left(-a_+ e^{-i\phi}+a_-e^{i\phi}\right)/2.
\end{equation}
By substituting this result in Eq.~(\ref{eq:EmF}) it can be seen that a variation of the trapping potential over the poloidal angle $\phi$ can be controlled by the choice of the dressing field's polarization. At $\rho=\rho_0$, the potential is determined by 
\begin{align}
    \left|B_+^\text{(t)}\right|^2=\frac{|a_+|^2+|a_-|^2}{4}-\frac{|a_+a_-|}{2}\cos(2\phi-\alpha),
\label{eq:poloidalpot}
\end{align}
where $\alpha=\text{arg}(a_+)-\text{arg}(a_-)$.
For a single circular component, i.e.\ $a_+=0$ or $a_-=0$, the potential minimum is independent of both $\theta$ and $\phi$, resulting in a flat potential over the toroidal surface.
For an elliptical field, $0<|a_+|\neq|a_-|>0$, the trap splits into two poloidal minima, i.e.\ it forms two rings at $\phi_{1,2}=\alpha/2(+\pi)$. For the extreme case of $|a_+|=|a_-|$, i.e.\ for linear polarization of $\textbf{B}_\mathrm{rf}^\text{(t)}$, the coupling at the minimum vanishes, leading to degenerate potentials for different $m_F$ and thus atom loss.


We are in particular interested in the case of forming rings at the top and bottom of the torus ($\phi_{1,2}=\pm\pi/2$). This occurs when the radial rf amplitude $a_r=(a_++a_-)/\sqrt{2}$ is smaller in magnitude than the vertical rf amplitude $a_z=(a_+-a_-)/\sqrt{2}$ and the corresponding real fields are $90^\circ$ out-of-phase, i.e.\ $\arg(a_z)=\arg(a_r)$ or $\arg(a_z)=\arg(a_r)\pm\pi$. Since the local static field in those rings is parallel to the $z$-direction, the trapping potential can be conveniently modulated by interference of $\textbf{B}_\mathrm{rf}^\text{(t)}$ with a multipole rf field, polarized in the $x,y$-plane, that oscillates at the same frequency. As shown below, the resulting modulation leads to the creation of state-dependent ring lattices.

\begin{figure}[t]
\centering
\includegraphics[width=.7 \columnwidth]{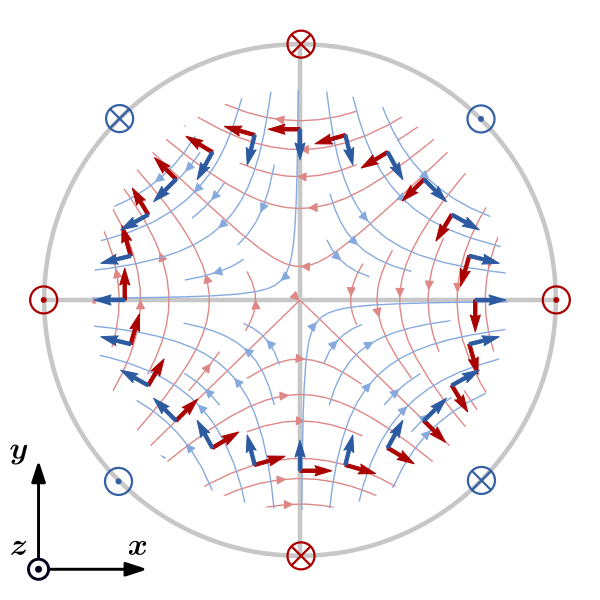}
\includegraphics[width=1\columnwidth]{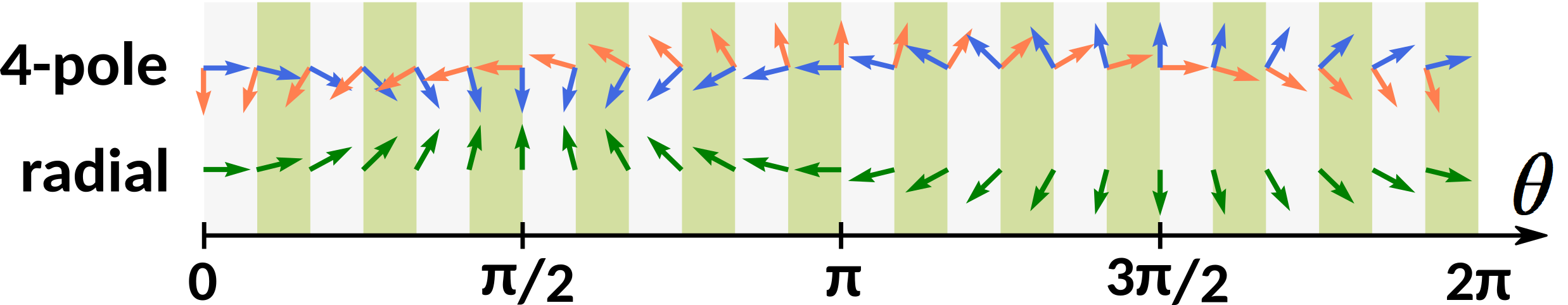}
\vspace{-0.3 cm}
\caption{\label{fig:quadrupole-basis}
Top panel: Orthogonal, linearly polarized, interior quadrupole fields ($n=2$) that can be driven with $90^\circ$-phase difference to generate circular polarization. The shown field lines are the leading order approximation (Eq.~\ref{eq:BxBy}) to the fields generated by two sets of four blue (red) infinitely long wires. The wires cross the plane at the locations depicted by the symbols $\otimes$ ($\odot$) for currents going into (out of) the page. The rotation offset angles are $\theta_0=0$ ($\theta_0=\pi/4 $) for the blue (red) quadrupole.
The bold arrows emphasize the local orthogonal field directions along a circular path, centered on the $z$-axis.
Bottom panel: Following the circular path in anticlockwise direction, the local field directions (blue and red arrows) rotate $n-1$ times clockwise, while the direction of a radial field pointing outwards (green arrows), rotates once in the opposite direction. The interference between these fields leads to $n$ minima and maxima. E.g., when the radial field is in phase with the blue multipole component, destructive interference occurs at $\theta=\pi/2$ and $\theta=3\pi/2$. Shifting the multipole phases by $90^\circ$ such that the phase of the red component aligns with that of the radial field leads to destructive interference at $\theta=\pi/4$ and $\theta=5\pi/4$. 
}\vspace{-0.3 cm}
\end{figure}

A linearly polarized, interior cylindrical multipole (m) field of order $n$ is described by
\begin{align}\label{eq:BxBy}
\textbf{B}_\text{rf}^\text{(m)}&=\mathcal{B}^\text{(m)}r^{n-1}\begin{pmatrix}\sin\left(\left(n-1\right)\theta-n\theta_0\right)\\
\cos\left(\left(n-1\right)\theta-n\theta_0\right)\\
0
\end{pmatrix},
\end{align}
where the offset angle $\theta_0$ describes a rotation about the $z$-axis and the moment $\mathcal{B}^\text{(m)}$ sets the field strength. The turning number of the field along a loop around the $z$ axis is given by $1-n$, and the expression in Eq.~\ref{eq:BxBy} can be viewed as including the field $\mathrm{B}_\text{rf}^\text{(c)}$ as a special case for $n=0$. For $n=1$, we obtain the homogeneous interior dipole field, $n=2$ describes a quadrupole field, etc. For $n>0$, the combination of two such fields of the same order, in particular the orthogonal cases for $\theta_0=0$ and $\theta_0=\pi/2n$, driven with individual rf amplitudes and phases, allows for the generation of a multipole field of arbitrary in-plane polarization. The example of two orthogonal, linearly polarized quadrupole fields is shown in Fig.~\ref{fig:quadrupole-basis}.

The generic (g) interior cylindrical multipole field can be expressed in terms of orthogonal circular components with moments $\mathcal{U}_\pm$, taking the form
\begin{align}
\boldsymbol{B}_\text{rf}^\text{(g)} &= \frac{\mathcal{U}_+}{\sqrt{2}}(re^{-i\theta})^{n-1}\begin{pmatrix}
        i \\
        1  \\
        0
     \end{pmatrix}  +
     \frac{\mathcal{U}_-}{\sqrt{2}}(re^{i\theta})^{n-1}\begin{pmatrix}
        i \\
        -1  \\
        0
     \end{pmatrix}.
\end{align}
Again, we neglect the radial dependence of this field in the vicinity of the traps ($r\approx r_0$) and approximate the local field amplitudes as $u_\pm=\mathcal{U}_\pm r_0^{(n-1)}$. The projection $B^\text{(g)}_+=\textbf{e}_+\cdot\mathbf{B}_\text{rf}^\text{(g)}$ of this field onto the relevant spherical component of the static field basis then leads to
\begin{equation}\label{eq:beta2pm}
B^\text{(g)}_+=-u_+\frac{1+\sin\phi}{2}e^{-in\theta}+u_-\frac{1-\sin\phi}{2}e^{in\theta}.
\end{equation}
At the top of the torus, i.e.\ for $\phi=+\pi/2$, the static field direction coincides with the setup's $z$-direction. Here, the amplitude $u_+$ leads to maximal coupling, while the contribution from $u_-$ vanishes. At the bottom of the torus, i.e.\ for $\phi=-\pi/2$, the static field direction is inverted, and we find the opposite situation.

\begin{figure}[b]
\centering
\includegraphics[width=0.9\columnwidth]{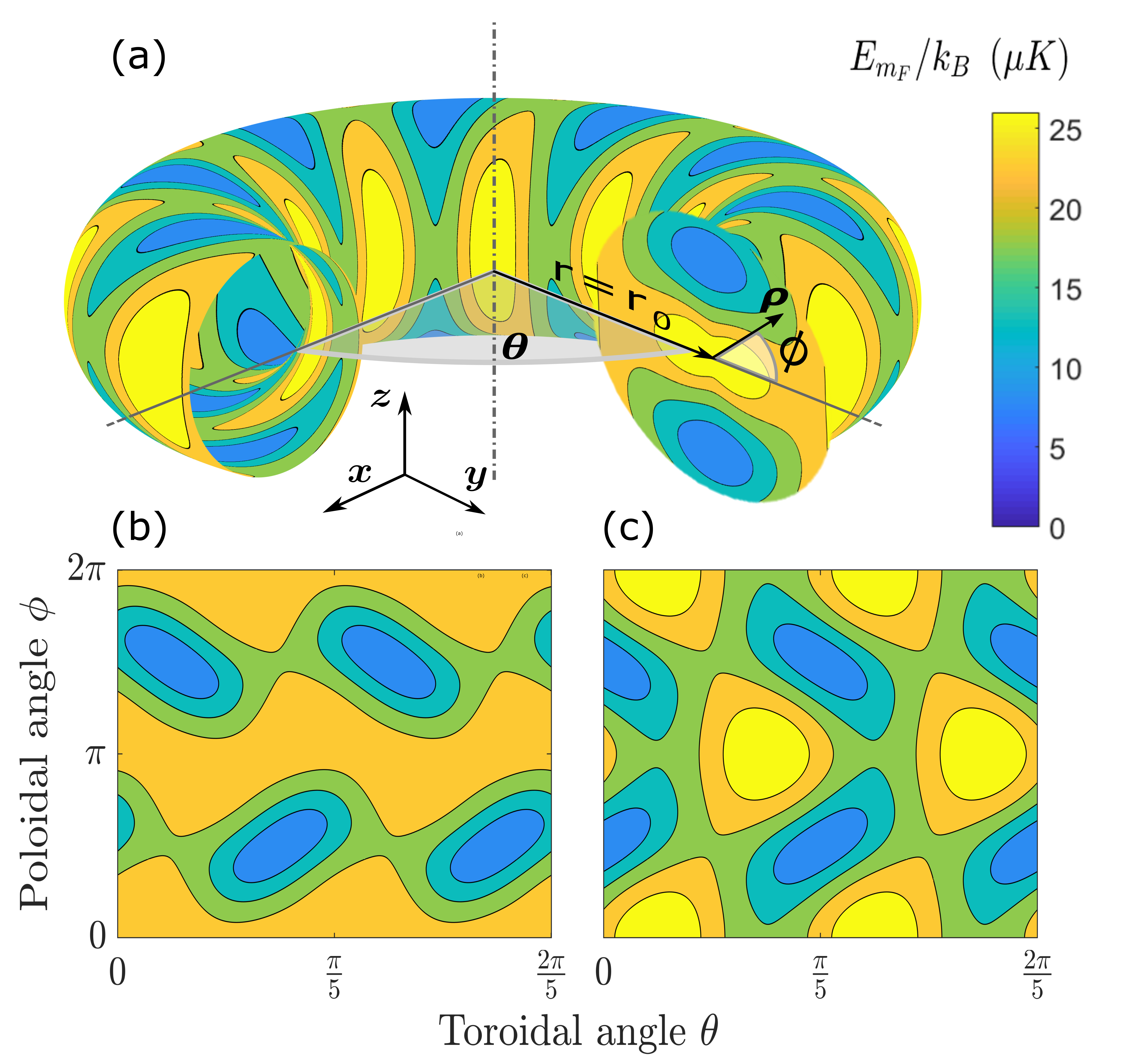}
\caption{\label{fig:trap_control} (a) Cuts through the approximated dressed potential (not to scale for realistic geometries). The potential landscape on the resonant torus surface is shown for $n=10$ traps. Potential minima are depicted as dark blue regions. On the right-hand side a cut through the 3D potential is shown for the $y,z$-plane. The assumed parameters are $m_Fg_F=1$, $a_r=0.9$~G, $a_z=1.5$~G, $|u_\pm|=0.3$~G. For a single internal spin state, the angular orientations and depths of the two lattices at the bottom and top of the torus can be controlled independently. This is shown for the potential on the resonant surface for $\varphi_\pm=0$ (b) and $\varphi_\pm=\pi/2$ (c).
For different species or spin states with different g-factors, independent potentials can be superimposed with some restrictions.}
\end{figure}

A ring lattice structure emerges from the combination of the toroidal field with a multipole field, because their amplitudes interfere. Assuming real amplitudes $a_\pm$ for the toroidal rf field and considering only the coupling at the top and the bottom of the torus, leads to
\begin{align}
\label{eq:beta2pm_ring}
|B_+|^2_{\phi=\pm\pi/2}=&|B_+^\text{(t)}+B_+^\text{(g)}|^2_{\phi=\pm\frac{\pi}{2}} \\
=&\frac{1}{2}a_r^2+|u_\pm|^2\pm \sqrt{2}a_r|u_\pm|\sin{\left(n \theta\mp \varphi_\pm\right)}\nonumber,
\end{align}
where we expressed $u_\pm=|u_\pm|e^{i\varphi_\pm}$. The result shows that for $|u_\pm|<|a_r|/\sqrt{2}$ the number of non-zero coupling strength minima and thus the number of traps formed around the rings is equal to the multipole order $n$, because it is given by the relative number of rf field rotations along the closed path [see Fig.~\ref{fig:quadrupole-basis}]. It also shows that the two circularly polarized quadrupole components control lattices in the top and bottom ring independently. The modulation depth and positioning of the forming lattices can be dynamically controlled via amplitudes and phases of the multipole rf fields (relative to the phase of the toroidal rf field).
The amplitudes enter with conjugate phases, which means that a single, linearly polarized multipole field that contains both amplitudes $u_\pm$ with a fixed phase difference will lead to counter-propagating lattices when this phase is ramped. 

Figure~\ref{fig:trap_control} provides a more comprehensive illustration of the potential landscape on the resonant surface for the case of two stacked rings, e.g., $a_+>0$, $a_-<0$, and both resonant fields $u_\pm$ present, generating $n=10$ traps in each ring. It can be seen that non-isotropic traps are formed that are not aligned with the toroidal and poloidal directions. In the harmonic approximation, this can be analysed by expressing the curvature tensor at the trap minima to determine aspect ratios and alignment. The squared trap frequencies are given by $\omega^2_{jk}=(\partial^2/\partial_j\partial_k)E_{m_F}/m$, where $m$ is the atomic mass of the trapped species.
For simplicity, we assume real $a_\pm$ again and make the assumption of only the locally dominant field $u_+$ or $u_-$ being present, although potential variations in one ring influence the confinement in the other ring to some degree.
To clarify radial, poloidal, and toroidal directions, and without loss of generality, we assume $\varphi_\pm=\pi/2$ and thus a potential minimum ($a_r>0$) or maximum ($a_r<0$) forming at $\theta=0$ and express the corresponding tensor in Cartesian coordinates $j,k\in\{x,y,z\}$   
\begin{align}
\left.\omega^2_{jk}\right|_{\phi=\pm\frac{\pi}{2}}=
    \frac{\gamma}{m}\left(
    \begin{array}{ccc}
        \frac{\delta^2+\frac{3}{2}a_r|u_\pm|}{\rho_0^2} & \frac{n a_z|u_\pm|}{r_0\rho_0} & 0  \\
        \frac{n a_z|u_\pm|}{r_0\rho_0} & \frac{n^2a_r|u_\pm|}{r_0^2} & 0 \\
        0 & 0 & 2\sqrt{2}q^2
    \end{array}
    \right),
\end{align}
using the definitions 
$\delta^2=(a_z^2-a_r^2-|u_\pm|^2)/\sqrt{2}$,
$\gamma=m_Fg_F\mu_B/(2\sqrt{2}B_0)$, and the effective field strength at the extremum $B_0=\frac{1}{2}|a_r-\sqrt{2}|u_\pm||$.
The comparison of diagonal and off-diagonal elements shows that the rotational sense of trap misalignment depends on the relative sign between amplitudes $a_z$ and $a_r$. In principle, the traps can be aligned by modulating the sign of $a_z$ to form a time averaged dressed potential \cite{lesanovsky_prl_2007, sherlock_pra_2011, navez_njp_2016}, which will make the averaged curvature tensor diagonal in these coordinates. The Hamiltonian at the trap centres will remain unmodulated, where $a_z$ does not enter the coupling strength [see Eq.~\ref{eq:beta2pm_ring}].

An important scenario is the trapping of atoms with different g-factors in the same rf dressed trap. Different atomic species may be controlled via different radio-frequencies~\cite{bentine_jphysB_2017}, and a particular case is a mixture or superposition of the same species in different hyperfine states. Here, the magnitude of the g-factor may be approximately equal but of opposite sign. In our description, a negative g-factor leads to a negative resonant dressing frequency. But the presence of any $a_\pm,u_\pm$ amplitudes with positive $\omega$ implies the presence of the corresponding amplitudes $a_\mp, u_\mp$ with negative $\omega$ and conjugated phase. The torus and ring forming potential in the poloidal direction is symmetric in this respect [see Eq.~(\ref{eq:poloidalpot})]. A trap formed at frequency $\omega$ also leads to a trap formed at frequency $-\omega$. But as it can be seen from Eq.~(\ref{eq:beta2pm_ring}), control of atoms via $u_+$ in the top ring and/or $u_-$ in the bottom ring imparts the same effects in the opposite ring on atoms with the opposite g-factor $g_F$. It should be noted, however, that changing the sign of $g_F$ also inverts the orientation of trap misalignment, because the negative resonant frequency $\omega_\text{rf}$ changes the sign of the anti-symmetric $a_z$ but not of the symmetric $a_r$ as $a_+$ and $a_-$ swap their values.
Within one ring, the two types of atoms can be controlled independently. Atoms in different spin states can be transported in opposite directions, thus making the configuration a candidate for guided Sagnac interferometer gyroscopes and other atomtronic applications. A lattice filled with atoms in one spin state could be immersed in a homogeneous ring of atoms in another spin state to couple different sites via phonons or study quantum friction of impurities in a Bose-Einstein condensate.

We briefly exemplify a few experimental considerations. With micro-fabricated trapping structures \cite{keil_jmodopt_2016} high field gradients can be achieved, e.g., $q=10$~T/m=$1/10$~G/$\mu$m. For $^{87}$Rb atoms in their electronic ground state, with total spin $F=2$, $g_F=1/2$, and $m_F=2$, a torus with $\rho_0\approx10~\mu$m forms for a dressing frequency $\omega_\text{rf}=700$~kHz. Assuming realistic amplitudes $a_r=0.9$~G, $a_z=1.1$~G, and $|u_+|=0.2$~G to form $n=10$ traps over a millimeter sized ring with $r_0=0.5$~mm leads to prolate traps with reasonable trap frequencies of $\omega_{xx}=2\pi\times989$~Hz, $\omega_{yy}=2\pi\times116$~Hz, and $\omega_{zz}=2\pi\times
2.3$~kHz
The coupling strength at the minima corresponds to $B_0=0.31$~G with a transition frequency to other dressed sub-levels, i.e.\ an rf Rabi frequency, of $\Omega=|g_F|\mu_B B_0/\hbar=2\pi\times227$~kHz. The Rabi frequency is sufficiently below the dressing frequency for the rotating wave approximation to be valid, but well above the highest trap frequency to avoid non-adiabatic atom loss. These parameters also allow for trap alignment via time-averaged, adiabatic potentials with a modulation of $a_z$ at tens of kHz that is sufficiently fast compared to atomic motion but slow enough for atomic spins to adiabatically follow.

Depending on the targeted geometry of the ring lattice, we have to revisit the approximations that were made to describe the underlying field geometries. We approximated the static field to be described by a local quadrupole field and assumed
the rf field amplitudes to be constant near the forming traps. The static field approximation is valid for $\rho_0\ll r_0,z_0$ where $z_0$ is a minimal distance from the field generating structures. For a thin torus with $\rho_0\ll r_0$, also the approximation for $B_+^\text{(t)}$ is valid, but  $B_+^\text{(g)}$ requires more careful examination, as $\mathbf{B}_\text{rf}^\text{(g)}$ scales with $r^{n-1}$. Here, the relative amplitude variation over the full resonant surface is given by
\begin{align}
    \eta=\left. \Delta \mathbf{B}_\text{rf}^\text{(g)}/\mathbf{B}_\text{rf}^\text{(g)}\right|_{r=r_0} =\frac{2\rho_0 (n-1)}{r_0},
\end{align}
which increases with the multipole order. For the numerical example above, we find $\eta=0.36$, which will already affect the shape of the dressed potential. However, for sufficiently low temperatures of the atomic cloud, the filled trap volumes will be confined to a much smaller range of radii $\Delta r<2\rho_0$, in particular when the traps are aligned by modulating the vertical rf field amplitude $a_z$.

\begin{figure}[bt]
\centering
\includegraphics[width=0.9\columnwidth]{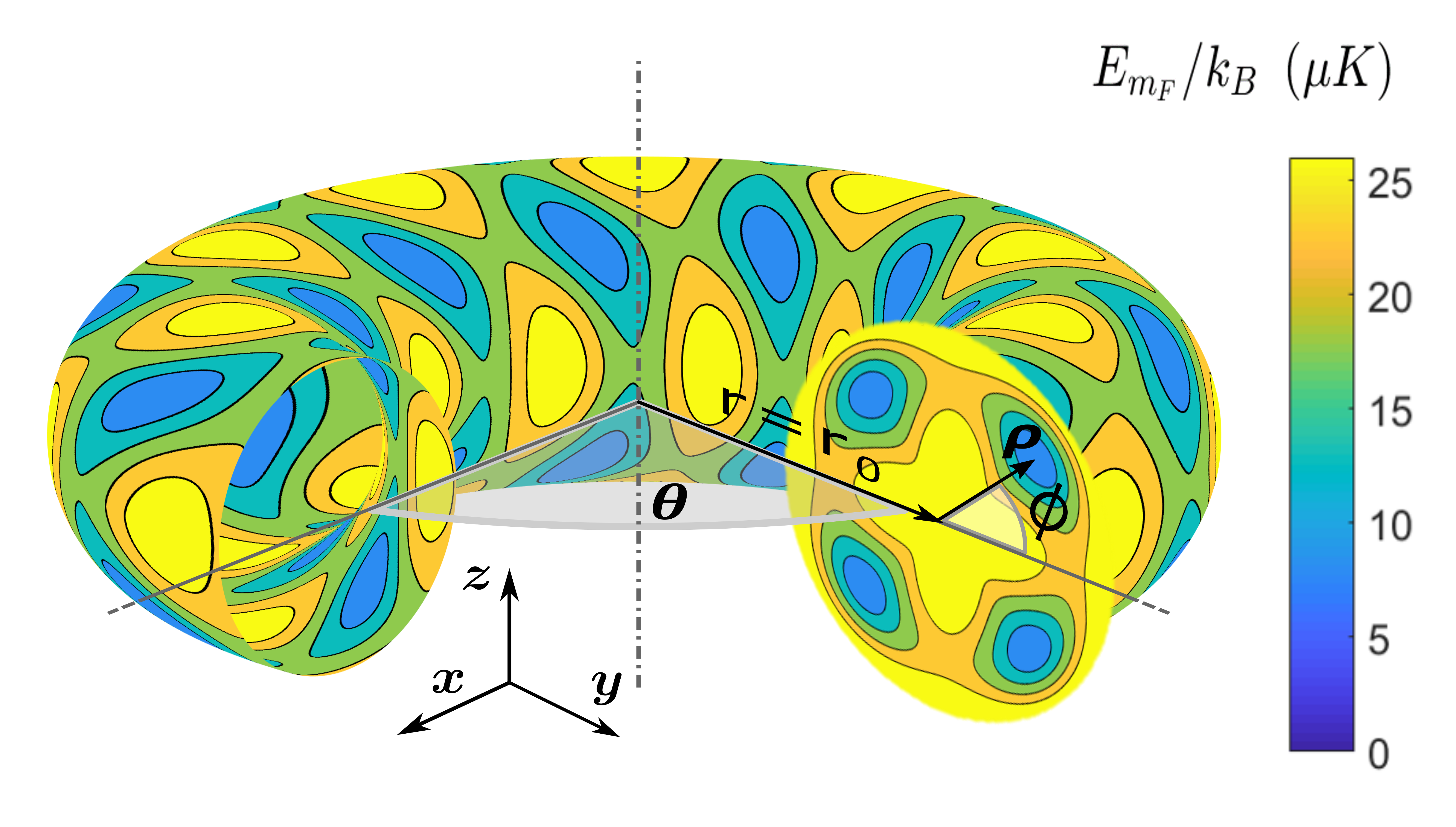}
\caption{Approximated dressed potential for multiple ring lattices, which form for static fields that are described by interior multipoles of order $l>1$ in the vicinity of a ring of zero field. The example shown uses the same dressing parameters as in Fig.~\ref{fig:trap_control}, but for a ring-shaped hexapole static field with $l=3$, which leads to $2(l-1)=4$ ring lattices at locations where the static field is aligned with the $z$-axis. \label{fig:mpole_static_field}}
\end{figure}

As an outlook, we can also consider higher-order multipoles to describe static fields for small $\rho$ by replacing $\phi\rightarrow(l-1)\phi$ in Eqs.~(\ref{eq:e0}-\ref{eq:e2}) together with the radial dependence $B_\text{dc}\rightarrow q_l\rho^{l-1}$. In this case, $2(l-1)$ rings occur, each with $n$ lattice sites, shown for $l=3$ in Fig.~\ref{fig:mpole_static_field}. But not all rings can be controlled independently. Tailored trap patterns could be generated by combining different rf dressing multipole fields of different orders.
Such patterns in combination with the ability to control these lattice sites in a state-dependent and dynamic fashion will create novel platforms for quantum simulation of interesting new physics. In general, the underlying principles are not restricted to the described toroidal geometries but allow for other combinations of inhomogeneous static fields with inhomogeneous dressing fields and a plethora of possible designs.

In summary, the scheme introduced in this work allows the generation of ring-shaped atom-trap lattices by only using static and rf magnetic fields.
Through the modulation of rf-dressed, toroidal potentials by rf-multipole fields (of order $n$) atom-trap lattices (with $n$ sites) can be created, that are state-dependent and also allow for dynamic, independent control over the potential landscapes for different atomic species or spin states. 
These features together with tight atom confinement that can be maintained also for large ring sizes make them a candidate for the realization of robust, guided Sagnac interferometer gyroscopes.

Since the components for the generation of the required fields for this scheme are also compatible with existing atom-chip technologies, integrated platforms can be designed and manufactured to produce such purely magnetic atom-trap lattices.
This will further extend the possibilities for compact quantum sensors that can be employed in the field and may enable novel quantum devices that can put fundamental physics questions to the test.
\\
\\

This work was funded by the Engineering and Physical Sciences Research Council (EPSRC), grant agreement EP/M013294/1 - UK Quantum Technology Hub for Sensors and Metrology.

\bibliographystyle{apsrev4-2}
\bibliography{rings}
\end{document}